\documentclass[a4paper, 11pt]{article}
\usepackage{amssymb,amsmath}
\usepackage[T1]{fontenc} 
\usepackage[utf8]{inputenc}
\usepackage{array}
\usepackage{booktabs} 	
\usepackage{caption}
\usepackage{graphicx}
\usepackage{subcaption}
\newcolumntype{R}{>{$}r<{$}}
\newcolumntype{L}{>{$}l<{$}}
\newcolumntype{M}{R@{${}\in{}$}L}
\usepackage{authblk}

\usepackage{hyperref}
\hypersetup{
    bookmarks=true,         
    unicode=false,          
    pdftoolbar=true,        
    pdfmenubar=true,        
    pdffitwindow=false,     
    pdfstartview={FitH},    
    pdfauthor={Kevin Multani},     
    colorlinks=true,       
    linkcolor=blue,          
    citecolor=red,        
}

\title{Measuring the neutron star compactness and
binding energy with supernova neutrinos}
\author[1,2]{A.\ Gallo Rosso\thanks{Corresponding author: andrea.gallorosso@gssi.it.}}
\author[1,2]{F.\ Vissani}
\author[3]{M.C.\ Volpe}
\affil[1]{Gran Sasso Science Institute,
Viale F.\ Crispi 7, L'Aquila, Italy}
\affil[2]{INFN,
Laboratori Nazionali del Gran Sasso,\\\hspace{\textwidth}
Via G. Acitelli, 22, Assergi, L'Aquila, Italy}
\affil[3]{Astro-Particule
et Cosmologie (APC),
CNRS UMR 7164, Universit\'e Denis Diderot,\\
10, rue Alice Domon et L\'eonie Duquet,
75205 Paris Cedex 13, France}
\date{}                     
\begin{document}
  \maketitle
  \begin{abstract}
We investigate the precision with which
a neutron star gravitational binding energy
can be measured through the supernova
neutrino signal, without assuming any prior
such as the energy
equipartition hypothesis, mean energies hierarchy 
or constraints on the pinching parameters that
characterize the neutrino spectra.
We consider water Cherenkov
detectors and prove
that combining inverse beta decay with
elastic scattering on electrons is sufficient
to reach $11 \%$ precision on the
neutron star gravitational binding energy
already with Super-Kamiokande.
The inclusion of neutral current events on
oxygen in the analysis does not improve the
precision significantly, due to theoretical 
uncertainties. We examine the possible impact on the conclusion 
of further theoretical input and of higher statistics.
We discuss the implications of our findings
on the properties of the newly formed neutron star,
in particular concerning the assessment of
the compactness or mass--radius relation.
\end{abstract}

\section*{Introduction} 

\addcontentsline{toc}{section}{Introduction} 

The importance of neutrino astronomy and
in particular of the observations of $\bar\nu_{\mathrm{e}}$
from SN~1987A \cite{k1,k2,Bionta:1987qt,
Bratton:1988ww, Alekseev:1988gp} is widely recognized.

A future  observation of a core collapse supernova
in the Milky Way, at a distance $3\div 10$ times
smaller than SN~1987A will have an impressive
scientific potential and it is eagerly awaited
by the scientific community. A special goal,
here discussed, is the possibility
of measuring directly the amount of gravitational
binding (potential) energy $\mathcal{E}_{\mathrm{
B}}$ of the newly formed neutron star, released
in neutrinos, thanks to the neutrino telescopes.
Besides those involved in SNEWS
\cite{Antonioli:2004zb}, new detectors
such as DUNE \cite{Ankowski:2016lab}, the
large scale JUNO \cite{j} and hopefully
Hyper-Kamiokande \cite{hk}
will also be operational.

The main sample of observed SN~1987A events
consists of $\bar\nu_{\mathrm{e}}$,  seen through
inverse beta decay (IBD), i.e.\
$\bar\nu_{\mathrm{e}}+\mathrm{p}\to \mathrm{e}^++\mathrm{n}$.
Under the hypothesis
that the energy is equally partitioned among
the six neutrinos species, the gravitational
energy was found to be $3  \times 10^{53}$
erg (at best-fit point and within errors)
in agreement with
expectations \cite{Vissani:2014doa}. 
The other neutrino and antineutrino species
are less easy to be seen. 
This requires in particular $\nu_{\mathrm{e}}$ sensitive
detectors such as those based  on liquid argon,
or on lead like HALO \cite{Vaananen:2011bf},
including carbon as for the large scale
scintillator detectors JUNO or oxygen 
in Super-Kamiokande \cite{sk} and the future
Hyper-Kamiokande. The non-electron
component of the neutrino fluxes could be
extracted through elastic scattering on
protons \cite{Beacom:2002hs}, and
the important role of elastic scattering
on electrons is discussed immediately below.

In principle, neutrino oscillations could
come to the rescue if the spectrum of $\bar\nu_{\mathrm{e}}$
probed by IBD would be composite, thereby offering
a chance to observe  the initial distributions of 
$\bar\nu_{\mathrm{e}}$ and also of  $\bar\nu_\mu$ and/or
$\bar\nu_\tau$. This happens in the simplest
case when the Mikheyev-Smirnov-Wolfenstein
effect takes place \cite{Dighe:1999bi}.
Moreover, the neutrino spectra
on Earth could also be composite due to the
neutrino self-interactions, or reach equilibrium
if fast conversion modes on short scales in the
supernova occur \cite{Sawyer:2015dsa}. 

On the other hand, astrophysical uncertainties
might undermine the possibility to extract information
on the emission spectra from the observations.
Ref.\ \cite{Minakata:2008nc} has pointed out that the
misreading of the pinching parameter implies that
the shape of the $\bar\nu_{\mathrm{e}}$
observed spectrum cannot
be used to learn about the emission spectra of
$\bar\nu_\mu$ and $\bar\nu_\tau$.  Moreover, it was
argued that the measurement of $\mathcal{E}_{\mathrm{B}}$
is compromised by the uncertainties in the shape
of the emitted neutrinos. This point was
originally made
by considering the IBD signal in Hyper-Kamiokande
\cite{hk}, but it applies also for Super-Kamiokande and JUNO,
that have a much smaller mass.

Is it possible to evade this conclusion, namely, is it
possible to measure the neutron star binding energy by
supernova neutrino observations?
Ref.\ \cite{Vaananen:2011bf} has argued that the range
of pinching parameters can be significantly constrained
by combining detection channels with different
energy thresholds and using other types of detectors. 
In this work, we will focus on 
neutrino measurements in Super-Kamiokande 
and analyze the possibility to
determine $\mathcal{E}_{\mathrm{B}}$ by exploiting 
other detection channels besides IBD in this detector.

We consider the fluences (i.e.\ the time integrated
spectra) for three different types of (anti)neutrinos,
namely,  $\nu_{\mathrm{e}},\bar\nu_{\mathrm{e}},\nu_x$
--- where $\nu_x$
means anyone of $\nu_\mu,\,\nu_\tau,\,\bar\nu_\mu$
and $\bar\nu_\tau$, supposed to have the same
initial distribution. For each one of these three
fluences, we introduce three parameters:  
1)~the emitted energy; 2)~the average energy;
3)~the parameter that accounts for deviations
from a thermal
distribution. Our work goes beyond existing analysis
in many respects. First of all we  vary these 9
parameters in wide ranges, to describe the effect
of the astrophysical uncertainties, without making
any assumption --- for example on energy
equipartition --- or fixing the pinching
parameter. Notice that energy equipartition is
often employed in this kind of analysis, while it
is not supported by current hydrodynamical
simulations of supernova explosions,
see e.g.\ \cite{Keil:2002in}.
We will give results under such hypothesis
as well, to show how they compare with the
available literature.

We consider  the signal due to IBD events as in
ref.\ \cite{Minakata:2008nc} and combine it with
elastic scattering  $\nu+\mathrm{e}^-\to \nu +
\mathrm{e}^-$ (ES) and neutral current events on
oxygen (OS). We demonstrate that the inclusion
of these
signals, and in particular of ES, allows the
measurement of $\mathcal{E}_{\mathrm{B}}$ with a
precision of about $10\%$. Our results also show
that a precision at a few percent level (and less)
is an achievable experimental goal. Finally, we
discuss the implications
of our findings, in particular for extracting the
compactness of the newly formed neutron star.


\section{Method of analysis}
\subsection{Neutrino properties}

We assume a supernova explosion at a distance
$D = 10$ kpc --- supposed to be known with a few percent
precision from astronomical observations.
We consider as true value of the total energy 
emitted in neutrinos $\mathcal{E}_{\mathrm{B}}^*$
the value $3\times 10^{53}$ erg, equally
distributed in fractions $\mathcal{E}_i^*$ among
the six species $(i = \nu_{\mathrm{e}},\,
\bar{\nu}_{\mathrm{e}},\,
\nu_x)$. These quantities are the standard ones used in the literature,
are not contradicted by astrophysical
simulations and are consistent with
SN~1987A data analyses
\cite{Lujan-Peschard:2014lta,Loredo:2001rx,Pagliaroli:2008ur}.
However, before proceeding, it should be emphasized that, for the
subsequent analysis, 
the equipartition ansatz  is not implemented as a constraint:
in other words,  
we do not assume to know which is the energy partition
and we perform a model independent analysis of the simulated data.

We investigate the neutrino and antineutrino
fluences, namely, their time-integrated fluxes. 
We parameterize the form of the fluence emitted at the source as 
suggested by \cite{Lujan-Peschard:2014lta,Tamborra:2012ac}
and, again, in agreement with numerical
simulations\footnote{In the following we put
$\hbar = c = k_B = 1$.}
\begin{equation}\label{eq:dFdE}
\mathcal{F}_i^0\left(E_{\nu}\right) = 
	\frac{\mathrm{d} F_i^0}{\mathrm{d} E_{\nu}} = 
	\frac{\mathcal{E}_i}{4\pi D^2}
	\frac{E_{\nu}^{\alpha_i}\:
	e^{-E_{\nu}/T_i}}{T_i^{\alpha_i +2 }\:
	\Gamma\left(\alpha_i+2\right)},
\end{equation}
with $i = \nu_{\mathrm{e}},\,
\bar{\nu}_{\mathrm{e}},\, \nu_x$ and where
$E_{\nu}$ is the neutrino energy, $\Gamma(x)$
is the Euler gamma function and $T_i =
\langle E_i\rangle / (\alpha_i + 1)$.
The true values of the parameters of the fluences
\eqref{eq:dFdE}, chosen for the simulation,  
are shown in table~\ref{tab:param}.
The assumptions on the
central values of 
the emitted energies $\mathcal{E}_i$ 
have been discussed previously.
Other remarks on the values listed in this table follows: 
\begin{enumerate}
\item The central values of the average energies
$\langle E_i\rangle$ do not contradict current theoretical simulations
and display a moderate hierarchy of values,
$\langle E_{\bar{\nu}_{\mathrm{e}}}\rangle/
\langle E_{\nu_x}\rangle =1.3$.
The mean energies come
from \cite{Lujan-Peschard:2014lta} and are
in agreement with what we know from
SN~1987A so far \cite{Vissani:2014doa}.
\item The central values of the pinching parameters $\alpha_i$
used for the true (simulated) spectra describe quasi-thermal
distributions.\footnote{This is a specific expected property of
time-integrated spectra. By contrast, the time-dependent fluxes are
expected to be strongly non-thermal especially at early times, and the 
corresponding pinching parameters can be much larger than
$\alpha_i=2.5$.}
The value $\alpha^*=2.5$ is the midpoint of the
	conservative interval $[1.5,\,3.5]$; 
	even 
	if its precise value is still
	unknown, for time integrated fluxes it is
	expected to be close to 2
	\cite{Vissani:2014doa},
	value that reproduces a Maxwell-Boltzmann distribution. 
\item The most important information from the table, especially for the following, 
are the conservative ranges that we employ to analyze the simulated data, with the aim to reconstruct
the (assumedly) true parameters. This statement applies for the 
9 astrophysical parameters (emitted energies, average energies, pinching parameters for three species) and also
for the oxygen cross section used for event detection (discussed below). 
\end{enumerate}
\begin{table*}[tbp]
	\centering
		\begin{tabular}{lMMM}
		\toprule
			& \multicolumn{2}{c}{$\nu_{\mathrm{e}}$}
			& \multicolumn{2}{c}{$\bar{\nu}_{\mathrm{e}}$}
			& \multicolumn{2}{c}{$\nu_x$}\\
			\midrule
			$\mathcal{E}_i^*\:[10^{53}\:\mathrm{erg}]$
			& 0.5 & [0.2,\,1] & 0.5 & [0.2,\,1]
			& 0.5 & [0.2,\,1]\\
			$\langle E_i^* \rangle\:[\mathrm{MeV}]$
			& 9.5 & [5,\,30] & 12 & [5,\,30]
			& 15.6 & [5,\,30]\\
			$\alpha_i^*$ & 2.5 & [1.5,\,3.5] & 2.5 & [1.5,\,3.5]
			& 2.5 & [1.5,\,3.5]\\
			\cmidrule{2-7}
			$\kappa^*$ &
			\multicolumn{2}{c}{~} &
			1 & [0.8,\,1.2] & \multicolumn{2}{c}{~}\\
			\bottomrule
		\end{tabular}
	\caption[True]{True parameter values assumed in the analysis
	and priors in which they can vary in the analysis.
	The quantities define the
	neutrino fluences --- namely, the time-intergrated fluxes
	--- given by eq.\ \eqref{eq:dFdE}.
	The first three rows describe the astrophysical parameters,
	while the fourth concerns the uncertainty in the OS
	cross section discussed later --- see eq.~\eqref{eq:LikOS}.}
 	\label{tab:param}
\end{table*}

The neutrino oscillation
mechanism is assumed to be described --- as a first
approximation --- by the
Mikheyev-Smirnov-Wolfenstein (MSW)
effect \cite{Wolfenstein:1977ue,Mikheev:1986gs}
under the hypothesis of normal hierarchy.
The fluences for electron neutrinos and antineutrinos
after oscillation become, in three flavors
\cite{Dighe:1999bi}
\begin{equation}\label{eq:fluIBD}
	\begin{cases}
		\mathcal{F}_{\nu_{\mathrm{e}}} &= \mathcal{F}_{\nu_x}^0\\
		\mathcal{F}_{\bar{\nu}_{\mathrm{e}}} &= 
		P_{\mathrm{e}}
		\cdot \mathcal{F}_{\bar{\nu}_{\mathrm{e}}}^0 + 
		(1-P_{\mathrm{e}}) \cdot\mathcal{F}_{\nu_x}^0,
	\end{cases}
\end{equation}
where the superscript 0 refers to the
emitted fluences at the source
and $P_{\mathrm{e}} = \left| U_{\mathrm{e}1} \right|^2 \approx
0.70$ \cite{Capozzi:2017ipn}; notice that the measured
$\nu_{\mathrm{e}}$ correspond to the emitted $\nu_x$.

\subsection{Expected events}

As possible channels of neutrino detection in
Super-Kamiokande we consider: 1)~inverse beta
decay (IBD); 2)~elastic scattering on electrons
(ES); 3)~quasi-elastic neutral current scattering
on \textsuperscript{16}O (OS).
We recall that: 1)~the first
reaction is
the one that gives the largest number of events
and the positrons are approximately
distributed in a isotropic manner;
2)~the second one provides
us events in a narrow forward cone of about
$20^\circ$ \cite{Nakahata:1998pz},
and this allows its identification with sufficient precision;
3) the last reaction, finally, produces measurable $\gamma$-ray
lines that will be treated simply
as a contribution to the total
number of events in a suitably
chosen low energy region,
discussed below.

In general, given a supernova
explosion characterized by the parameters
specified
above, the number of expected events for the
reaction $j$ generated by the neutrino species
$i$ can be expressed as
\begin{equation}
	\mathrm{N}_{i,j} = N_{T,\,j} 
	\int_{E_{\nu,\mathrm{min}}}^{\infty}
	\mathrm{d}{E_{\nu}}\:
	\mathcal{F}_{i}\left(E_{\nu}\right)
	\sigma_{i,j}\left(E_{\nu}\right),
\end{equation}
where $E_{\nu}$ is the neutrino energy,
$N_{T,\, j}$ is the number of targets for
the process $j$ and $\sigma_{i,
j}\left(E_{\nu} \right)$ is the energy dependent
cross section for a given reaction and species.
The integral goes from the minimum energy
$E_{\nu,\mathrm{min}}$
to 300 MeV, which for all purposes
is the same as infinity.
For IBD and ES events in Super-Kamiokande
we assume a threshold
on the recoiling $\mathrm{e}^{\pm}$
of 5 MeV \cite{Beacom:1998ya,Smy:2010zza}.
Concerning the OS signal, it is expected
to be within a window of $4\div 9\:\mathrm{MeV}$,
obtained combining the
window covered by the expected
gamma lines ($\approx 5.3\div 7.3$ MeV)
\cite{Langanke:1995he} with
the energy resolution of the detector
($\approx 1.1\div 1.3$ MeV)
\cite{Fukuda:2002uc}.
In this region, it cannot be disentangled from
the (many more) IBD and ES background events.
Thus, we are bound to consider the
sum of the three contributions. This constitutes
the \emph{neutral current low energy region} (NCR)
that includes the OS signal.
The number of expected events for each
reaction is reported in table~\ref{tab:estratti}.

\begin{table}[tp]
	\centering
	\begin{tabular}{lccccr}
		\toprule
 		& $\nu _e$ & $\bar{\nu }_e$ & $\nu_{x}$
 		 &  sum & extracted \\
 		\midrule
 		IBD 	& ---	& 2900	& 1672	& 4572  & 4565\\
 		ES 		& 14.7 	& 24.7 	& 187 	& 226 	& 237 \\
 		\cmidrule{1-6}
 		NCR		& 11.6 	& 345 	& 204	& 561 	& 554\\
 		OS sig.\  & 0.53	 	& 2.04 	& 43.0 	& 45.5 	\\
 		IBD bkg.\ & ---		& 324 	& 77.8 	& 401 	\\
 		ES	bkg.\ & 11.0 	& 19.2 	& 83.3 	& 114 	\\
 		\bottomrule
	\end{tabular}
	\caption{Number of expected events in Super-Kamiokande,
	divided by the contribution given by each neutrino
	species.   We use the notation 
	$\nu_x = \nu_{\mu}
	+\bar{\nu}_{\mu} + \nu_{\tau} + \bar{\nu}_{\tau}$.
	The results correspond to the true parameters'
	values used in the analysis, and reported in
	table~\protect\ref{tab:param}.
  	Remarks on NCR and OS events can be
  	found in the text.
	The last column refers to the values
	extracted for
	the statistical analysis presented
	in this paper.}
 	\label{tab:estratti}
\end{table}

\subsection{Statistical procedure}

As likelihoods for IBD and ES events we use
a standard binned form with the same bin widths
as in \cite{Minakata:2008nc}. Concerning the OS
events we use a Gaussian function, with a caveat.
For the OS cross section we assume the
analytic form
reported in \cite{Beacom:1998ya}
\begin{equation}
	\sigma_{\mathrm{OS}}(E_{\nu}) \approx
	\sigma_0
	\left(E_{\nu}/\mathrm{MeV} - 15\right)^4
	\quad\text{for}\quad E_{\nu}
	\ge 15\:\mathrm{MeV},
	\label{eq:CsecParam}
\end{equation}
where $\sigma_0 =
4.21\times 10^{-22}\:\mathrm{fm}^2$
for neutrinos and $\sigma_0 = 3.33\times
10^{-22}\:\mathrm{fm}^2$ for
antineutrinos.\footnote{Their values can
be inferred
from table I of \cite{Langanke:1995he},
from the branching ratios
that explicitly have a $\gamma$-ray in
the final state}.

In the literature, quasi-elastic neutral current
cross section on oxygen can be found as
computed by many models, e.g.\ 
\cite{Haxton:1987kc,Kolbe:1992xu,
Langanke:1995he,Kolbe:2002gk}.
Comparing the calculations in the literature,
the uncertainty seems to
be of the order of several 10\%. 
Therefore, we assume a optimistic but not
unrealistic value of the uncertainty, namely 10\%,
that could be
hopefully reached in the future.

Thus, we introduce a tenth parameter, $\kappa$,
as a multiplicative constant for the whole cross
section (identical for neutrinos and
antineutrinos) in order to parametrize the 
systematic uncertainty.
It varies according
to a Gaussian of mean $\kappa^* = 1$ and
standard deviation $\sigma_{\kappa} = 0.1$,
in a prior $[0.8,1.2]$, see table~\ref{tab:param}.
The NCR likelihood becomes
\begin{equation}\label{eq:LikOS}
	\mathcal{L}_{\mathrm{NCR}}\left(\text{param.}\right)
	\propto\text{exp}\left[
	-\frac{\left(n_{\mathrm{NCR}} -
	\text{N}_{\mathrm{NCR}}\right)^2}{2\text{N}_{\mathrm{NCR}}} -
	\frac{\left(\kappa -1 \right)^2}
	{2\sigma_\kappa^2}\right],
\end{equation}
where $\text{N}_{\mathrm{NCR}}$ is the number of
expected events as a function of the parameters
and $n_{\mathrm{NCR}}$ is the extracted one
(see third row of table~\ref{tab:estratti}).
It is useful to anticipate that
our numerical analysis shows that, despite
this optimistic assumption on $\kappa$,
the inclusion of the NCR has only
a minor impact on the conclusion. Moreover,
it depends only weakly on the 
(supposed) value of $\kappa$.
In fact, as shown in table~\ref{tab:estratti},
the amount of NCR events that can
be affected by a modification in the cross
section is really small if compared to the
IBD+ES related background. For example, increasing
$\kappa$ from 1.0 to 1.2 rises the number
of OS events from 45.5 to 54.6, leading
to an almost identical result.

In order to quantify the relevance of the various reactions,
we perform three different analyses: 1)~we start with
IBD events alone; 2)~then we add the ES information;
3)~and eventually we include the NCR.

In the analyses we assume full tagging efficiency for the
ES process. This assumption is justified as follows.
The directional discrimination on the ES events
allows us to reduce the IBD background to the $20^\circ$ cone
in which it overlaps to the ES signal --- the 3\% of the
solid angle. Then, 20\% of those IBD events can be
rejected with the neutron tagging and a further 20\%
requiring a visible energy lower than
30 MeV \cite{Pagliaroli:2009qy}, already in the current configuration of 
Super-Kamiokande. Putting all together
we end up with $\sim 100$ IBD-beam related background events,
whose statistical variation is $\sim 10$, smaller than the
one of the ES events.
Moreover, gadolinium doping
\cite{Beacom:2003nk} can 
provide an important improvement on neutron
tagging \cite{Laha:2013hva}. According to current
data and simulations, the IBD-beam related background events 
would lower to 20\%; in fact, the efficiencies of neutron tagging
is $\sim 0.9$ and 
the neutron reconstruction efficiency is again $\sim 0.9$
\cite{Labarga:2016nbw}.

The statistical procedure follows a Monte
Carlo approach:
$n-$dimensional random points are extracted in the
region described by the priors listed in
table~\ref{tab:param}. Then, each point $P$
is accepted within a certain confidence level (CL) if its
likelihood satisfies the relation
\begin{equation}
	\log\mathcal{L}\left(P\right)\ge
	\log\mathcal{L}_{max}
	- \frac{A}{2},
\end{equation}
where $\mathcal{L}_{max}$ is the likelihood
maximum inside the prior and $A$
is defined with an integral of a
chi-square distribution with
$N_{\mathrm{dof}}$ degrees of freedom
\begin{equation}
	\int_0^{A} \chi^2(N_{\mathrm{dof}}; z)\,\mathrm{d}{z}
	= \mathrm{CL}
\end{equation}
For instance, considering $\mathrm{CL} = 0.9973$, namely
$3\sigma$, we find $A = 20.0621$, $23.5746$, $26.9011$
for $N_{\mathrm{dof}} = 6,\,8,\,10$ respectively.
This routine has been validated running the algorithm
on the analysis \cite{Minakata:2008nc} and obtaining
identical results.


\section{Results}

\begin{figure}	
	\centering
	\begin{subfigure}[t]{0.48\textwidth}
		\centering
		\includegraphics[width=\textwidth]{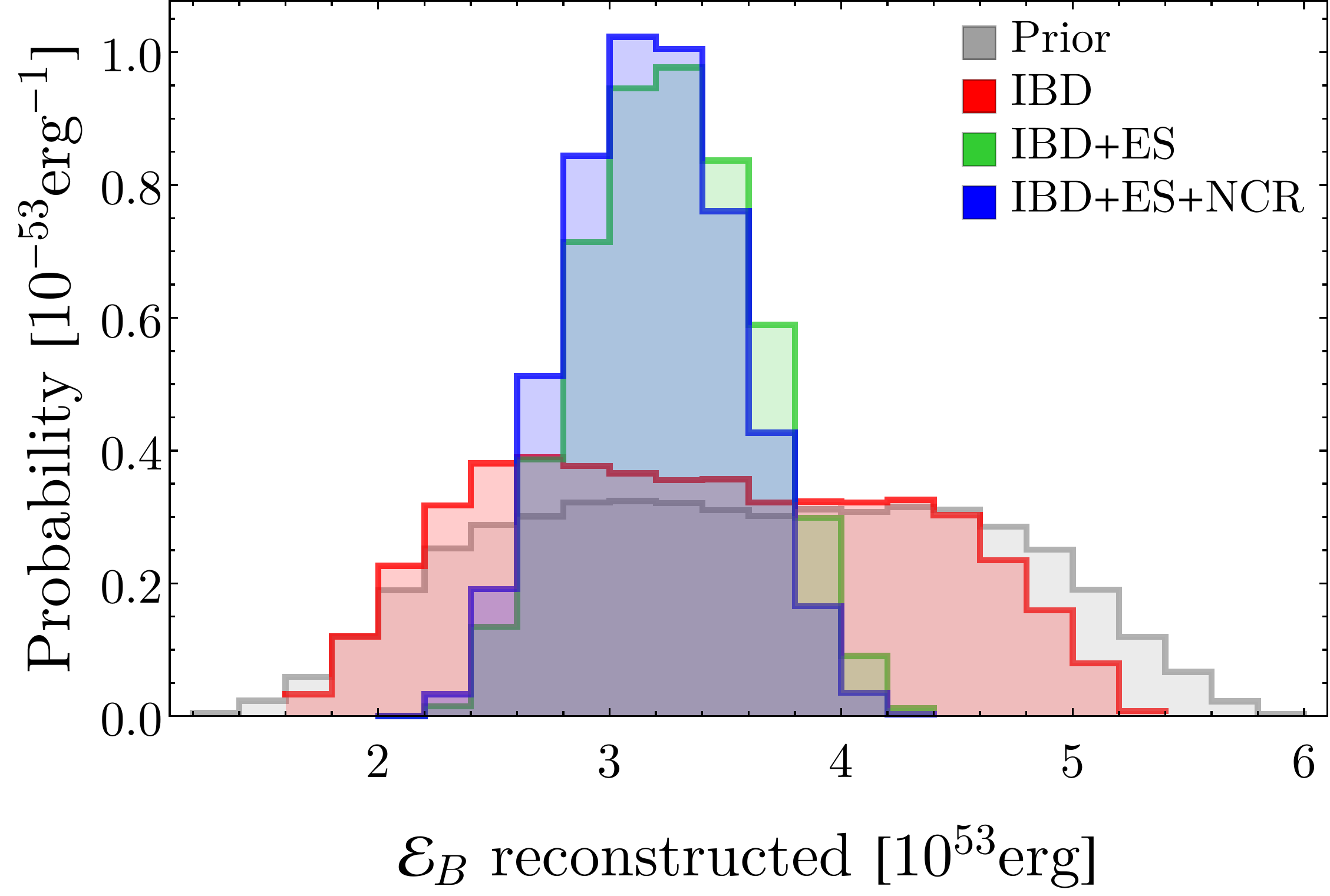}
		\caption{Standard}\label{fig:ris1}		
	\end{subfigure}
	\begin{subfigure}[t]{0.48\textwidth}
		\centering
		\includegraphics[width=\textwidth]{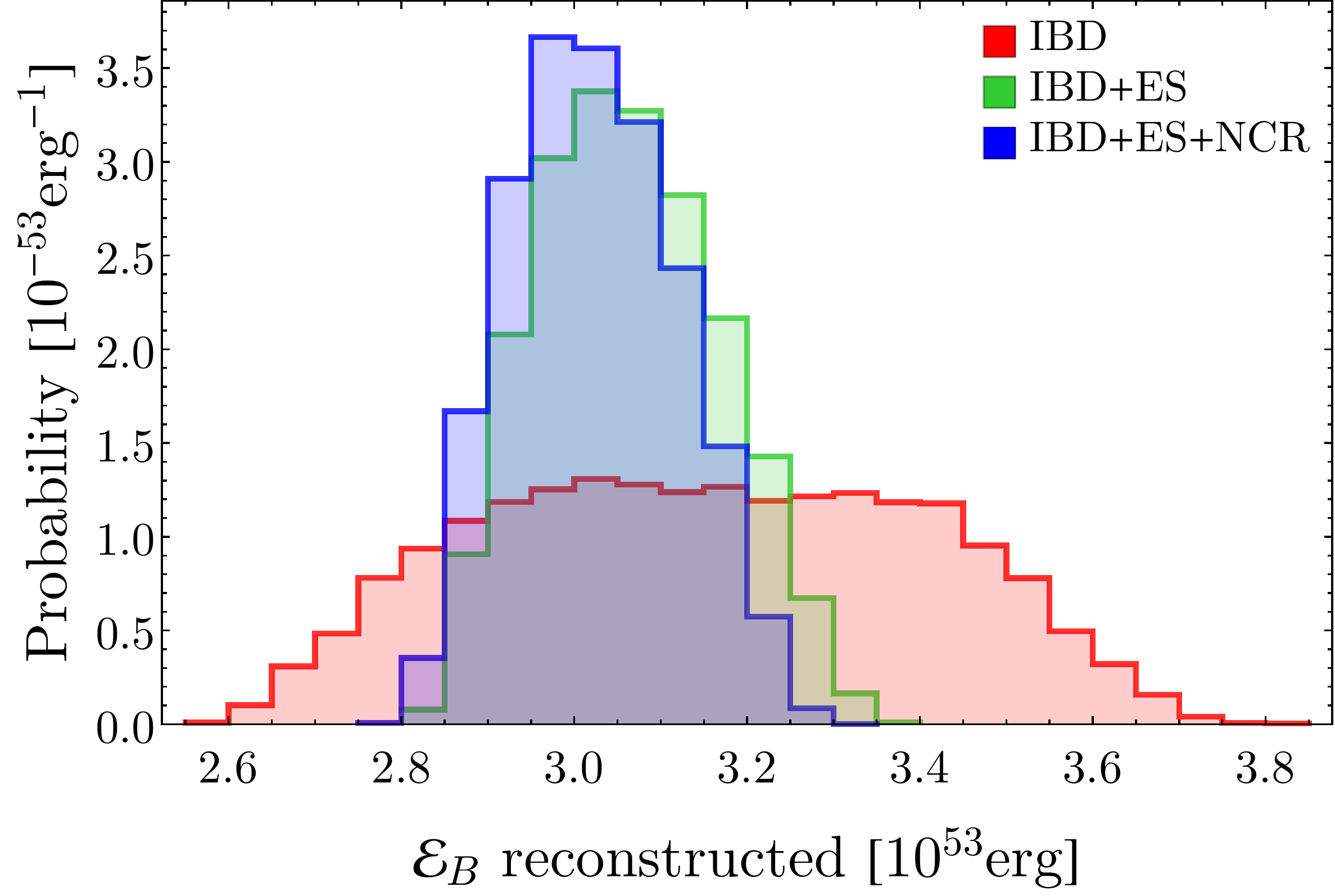}
		\caption{Equipartition}\label{fig:ris2}
	\end{subfigure}
	\caption{Reconstructed total energies $\mathcal{E}_{\mathrm{B}}$
in the analysis without any prior (\subref{fig:ris1}) and
with implemented energy equipartition among 
species (\subref{fig:ris2}). Different colors represent different
amount of information. The prior gray distribution
in (\subref{fig:ris1}) is obtained
extracting random uniformly $\mathcal{E}_{\nu_{\mathrm{e}}}$,
$\mathcal{E}_{\bar{\nu}_{\mathrm{e}}}$, $\mathcal{E}_{\nu_x}$ in
$[0.2,\,1]\times 10^{53}$ erg. Each distribution contains
30k extracted points and is normalized to 1.
Notice that the horizontal
scales in the two cases are very different.}\label{fig:risultati}
\end{figure}

For each extracted point $P$ the total energy can
be reconstructed as
\begin{equation}
	\mathcal{E}_{\mathrm{B},P} =
	\mathcal{E}_{\nu_{\mathrm{e}},P} +
	\mathcal{E}_{\bar{\nu}_{\mathrm{e}},P}
	+ 4 \mathcal{E}_{\nu_x,P},
\end{equation}
with a caveat for the analysis on IBD events alone:
since $\mathcal{E}_{\nu_{\mathrm{e}},P}$
cannot be measured, it is taken
randomly inside the prior. The results are gathered
in a histogram and shown in figure~\ref{fig:ris1}.

The distribution of the IBD-only analysis is quite similar to
the prior distribution, and this implies that it is not
possible to measure the total
energy by using IBD events only. This negative conclusion
fully agrees with the findings by \cite{Minakata:2008nc}.
When we add the ES events in the
analysis, the distribution changes,
as shown by the peak in figure~\ref{fig:ris1}:
the inclusion of the ES scattering
events allows us to measure the total energy with
Super-Kamiokande.
This is a new result that changes
qualitatively the conclusion concerning the relevance of
water Cherenkov detectors.

\begin{figure}	
	\centering
	\begin{subfigure}[t]{0.48\textwidth}
		\centering
		\includegraphics[width=\textwidth]{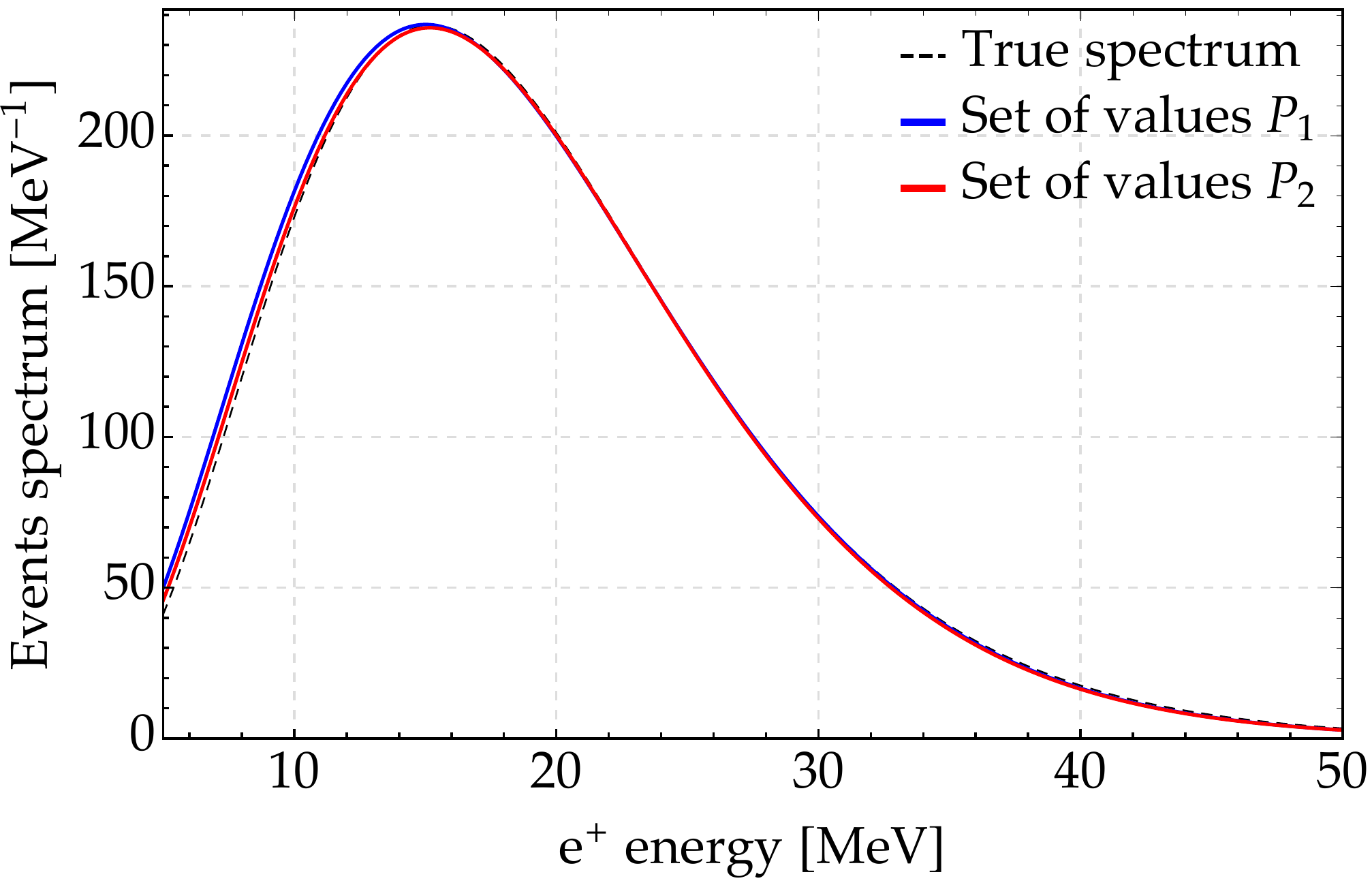}
		\caption{IBD}\label{fig:spe1}		
	\end{subfigure}
	\begin{subfigure}[t]{0.48\textwidth}
		\centering
		\includegraphics[width=\textwidth]{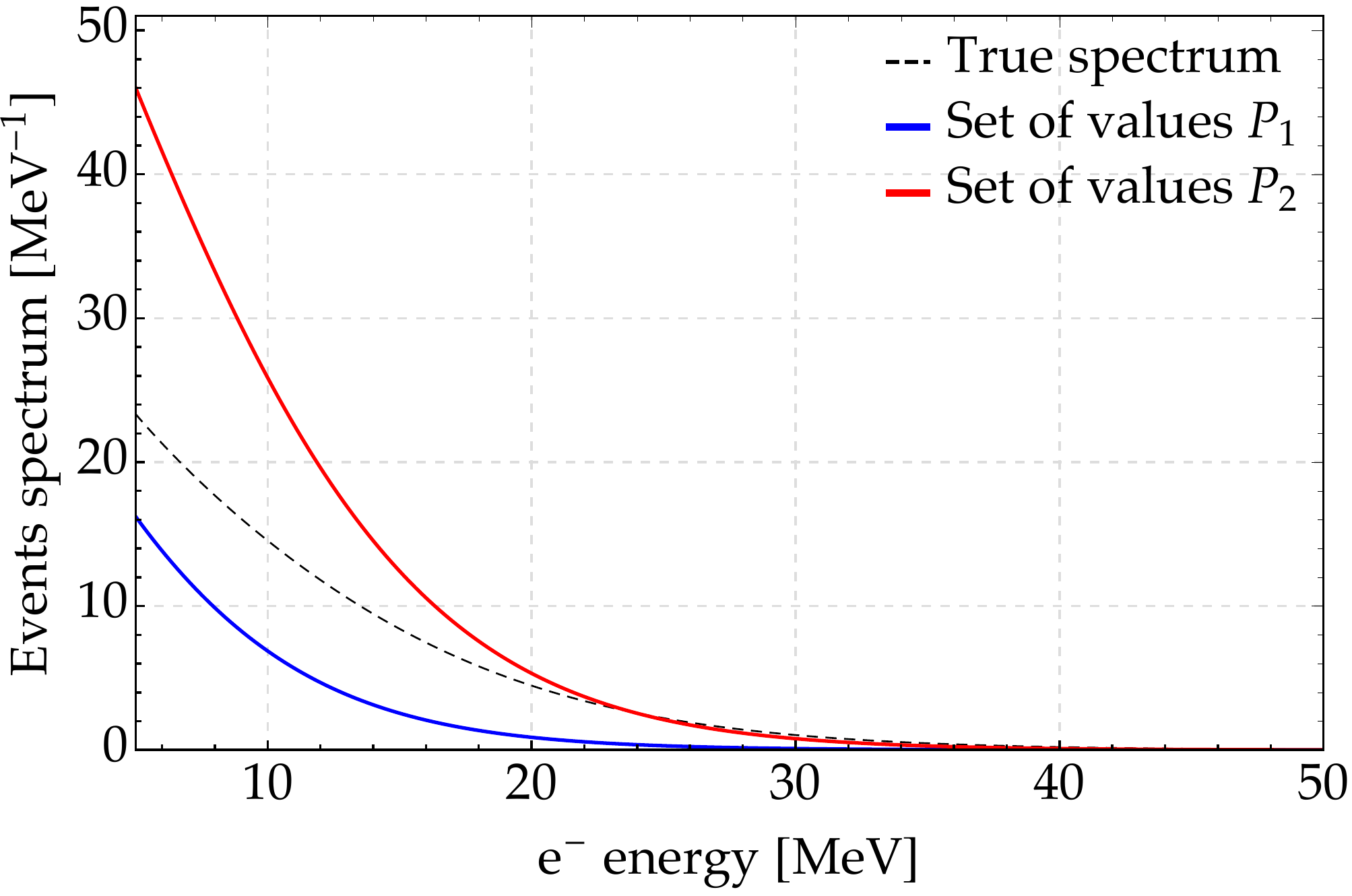}
		\caption{ES
		($\bar{\nu}_{\mathrm{e}}+\nu_x$)}
		\label{fig:spe2}
	\end{subfigure}
	\caption{IBD spectrum (\subref{fig:spe1}) and contribution by
$\bar{\nu}_{\mathrm{e}}$, $\nu_x$
to the ES spectrum (\subref{fig:spe2})
for two set of points, characterized by very different
values of $\mathcal{E}_{\bar{\nu}_{\mathrm{e}}}+
4\mathcal{E}_{\nu_x}$.
The first one is $P_1$ =
($\mathcal{E}_{\bar{\nu}_{\mathrm{e}}}$, $\mathcal{E}_{\nu_x}$,
$\langle E_{\bar{\nu}_{\mathrm{e}}}\rangle$,
$\langle E_{\nu_x}\rangle$,
$\alpha_{\bar{\nu}_{\mathrm{e}}}$, $\alpha_{\nu_x}$) =
($0.665 \times 10^{53}$ erg, $0.2\times 10^{53}$ erg,
12.75 MeV, 9.25 MeV, 2.075, 2.155) and the other one is
$P_2$ = ($0.294 \times 10^{53}$ erg,
$1 \times 10^{53}$ erg, 13.45 MeV, 11.9 MeV, 2.075, 2.155).
The sums $\mathcal{E}_{\bar{\nu}_{\mathrm{e}}}+
4\mathcal{E}_{\nu_x}$
are 1.5 and $4.3\times 10^{53}$ erg respectively.
Both points are accepted in $3\sigma$ CL IBD-only analysis
and they can be discriminated only adding the ES reaction.}\label{fig:spettri}
\end{figure}

This result can be understood
looking at figure~\ref{fig:spettri}:
two accepted points at $3\sigma$ CL,
with very different
$\mathcal{E}_{\bar{\nu}_{\mathrm{e}}}+
4 \mathcal{E}_{\nu_x}$, have similar IBD spectrum --- the same within
fluctuations --- and can be distinguished through the
contribution they give to the ES events.
Extracting mean and standard deviation from
histogram(s)~\ref{fig:ris1} we get
\begin{equation}
	\frac{\mathcal{E}_{\mathrm{B}}}{10^{53}\:\mathrm{erg}}
	\xrightarrow{\mathrm{IBD}}
	3.4 \pm 0.9 \xrightarrow{+\mathrm{ES}}
	3.3 \pm 0.4 \xrightarrow{+\mathrm{NCR}}
	3.2 \pm 0.4.
\end{equation}
We conclude that, for the combined
three-channels analysis,
the total energy can be
reconstructed within an accuracy of about 10\% 
(more precisely, $\approx 11\%$). 
This is the key result of
our analysis.

\paragraph{Remarks}
Often, analyses of simulated data that aim to
forecast the physics reach of a future supernova
assume equipartition hypothesis for granted.
The results obtained if we make this assumption
are shown in figure~\ref{fig:ris2}: 
The three channel combined analysis gives
$\mathcal{E}_{\mathrm{B}} = (3.02 \pm 0.09)
\times 10^{53}\:\mathrm{erg}$ with an accuracy
of $\approx 3\%$. However, the equipartition 
hypothesis is expected to be reliable only within a factor of two
\cite{Keil:2002in,Keil:2003sw,Raffelt:2005fb}.
In conclusion, the 10\% precision  should be considered as reference 
value  with current theoretical understanding,
since this is model-independent.

The inclusion of the neutral current events
in the NCR region yields a result consistent
with the IBD+ES analysis, however it does
not improve the determination significantly.
If the number of  beam-related background
events due to IBD reaction 
decreases to 20\%, thanks to the tagging
allowed by gadolinium doping, 
the accuracy for the combined
three-channels analysis remains almost the same,
namely changing from 11.0\% to 10.7\%.
This conclusion does not depend on the uncertainty
on the neutral current cross section. In fact,
even assuming an ideal scenario in which the
IBD background is 20\% and the OS cross section
is known perfectly, the resolution becomes
$\approx 10.3\%$. In other
words, the improvement due to the inclusion
of the NCR is very small.

Therefore, we conclude that, in order to measure
the neutron star binding energy with
Super-Kamiokande or similar neutrino telescopes,
it is very important to observe and analyze the
events due  elastic scattering of neutrinos and
electron, even more than the neutral current
events on oxygen. The fact that we include 
these events in our analysis explains the
difference with the conclusion of
ref.\ \cite{Minakata:2008nc},
that was based only on IBD events. 


\section{Discussion}
The precise measurement of the gravitational
binding energy in a future galactic supernova
explosion is of interest for astrophysics,
particle and nuclear
physics. In supernova theory, the \textit{delayed
neutrino-heating mechanism},
elaborated by Wilson \cite{Wilson:1971}
and by Bethe and Wilson \cite{Bethe:1984ux}, 
is currently thought to be the mechanism producing
the blow of most of supernovae type II and Ib/c. 
Close to giving successful explosions \cite{Muller:2017hht},
three-dimensional simulations include realistic neutrino
transport, nuclear networks, hydrodynamic instabilities,
convection and turbulence. 
The precise measurement of the gravitational binding
energy in a future galactic explosion would provide key
confirmation of the current paradigm of the delayed
neutrino-heating mechanism. From the particle
physics point of view,
improved limits would be gathered on neutrino properties
related to new physics, such as  
neutrino decay, the neutrino magnetic moment or the
existence of sterile neutrinos.

Interestingly, the determination of $\mathcal{E}_{\mathrm{B}}$ 
would give the star compactness
\begin{equation}\label{e:beta}
\beta = \frac{G M}{R\, c^2},
\end{equation}
where $G$ is the gravitational constant, $M$ and $R$
are the gravitational mass and the radius of the newly
 formed neutron star.
From the fit to the neutron star binding energies, for
a large set of equations-of-state (EOS) that permit
maximum masses larger than 1.65 $M_{\odot}$
as a function of $\beta$, one
has\footnote{Notice that this relation, including some
dependence on the radius, improves on a previous relation
by Lattimer and Yahil, namely
$\mathcal{E}_{\mathrm{B}}
\approx 0.084 \, (M/M_{\odot})^2M_{\odot} c^2$
that is also accurate to $\sim 10 \%$
\cite{Lattimer:1989zz}.} \cite{Lattimer:2006xb}
\begin{equation}\label{e:beta-eb}
\frac{\mathcal{E}_{\mathrm{B}}}{M c^2}
\approx \frac{\left(0.60 \pm 0.05\right)
\beta}{1 - \beta/2 },
\end{equation}
where $\mathcal{E}_{\mathrm{B}} = N_b m_b c^2
- M c^2 = M_b c^2 - M c^2$
is the gravitational binding energy, $M_b$ is the baryonic mass
corresponding to $N_b$ baryons of mass $m_b$.
The latter can be taken as the mass of a neutron or a proton
or $^{56}\mathrm{Fe} / 56$ for a white-dwarf iron core.

The baryonic mass can be determined from simulations by
considering the amount of mass inside the shock radius about
0.5 seconds after bounce, at the outer edge of the
\textsuperscript{56}Ni region and subtracting the amount
of nickel in the supernova ejecta. 
After inserting the expression for
$\mathcal{E}_{\mathrm{B}}$, the
$\beta-\mathcal{E}_{\mathrm{B}}$
relation \eqref{e:beta-eb} reads
\begin{equation}\label{e:beta-ebb}
\beta = \frac{\mathcal{E}_{\mathrm{B}}}{0.6\,M_b c^2
- 0.1\,\mathcal{E}_{\mathrm{B}}}.
\end{equation}
For SN~1987A, using eq.~\eqref{e:beta-eb}
and considering that the
$M_b = 1.733\, M_{\odot}$ and
$M = 1.53\, M_{\odot}$  \cite{Bethe:1995hv} 
one gets $\beta = 0.194$ and $R = 11.5$ km.

Relation \eqref{e:beta}
can also be used to get $R-M$ relation
\begin{equation}\label{e:r-m}
R = \frac{0.6\, G M^2}{\mathcal{E}_{\mathrm{B}}}
+ \frac{r_{\mathrm{s}}}{4},
\end{equation}
where we have introduced the Schwarzschild
radius $r_{\mathrm{s}} = 2 G M /c^2$.
Alternatively, one can obtain
the $M-R$ relation
\begin{equation}\label{e:m-r}
M = \sqrt{\frac{\mathcal{E}_{\mathrm{B}} R}{0.6\, G}}
\left[\sqrt{1 + \epsilon^2} - \epsilon
\right],
\end{equation}
where $\epsilon$ is defined as:
\begin{equation}
\epsilon = \frac{1}{4}\sqrt{
\frac{\mathcal{E}_{\mathrm{B}} G}{0.6\, R\, c^4}}.
\end{equation}
The main relativistic effect is given by the negative
term $\propto \epsilon$, namely $0.42\,\mathcal{E}_{\mathrm{B}}/c^2$.

\begin{figure}	
	\centering
	\begin{subfigure}[t]{0.48\textwidth}
		\centering
		\includegraphics[width=\textwidth]{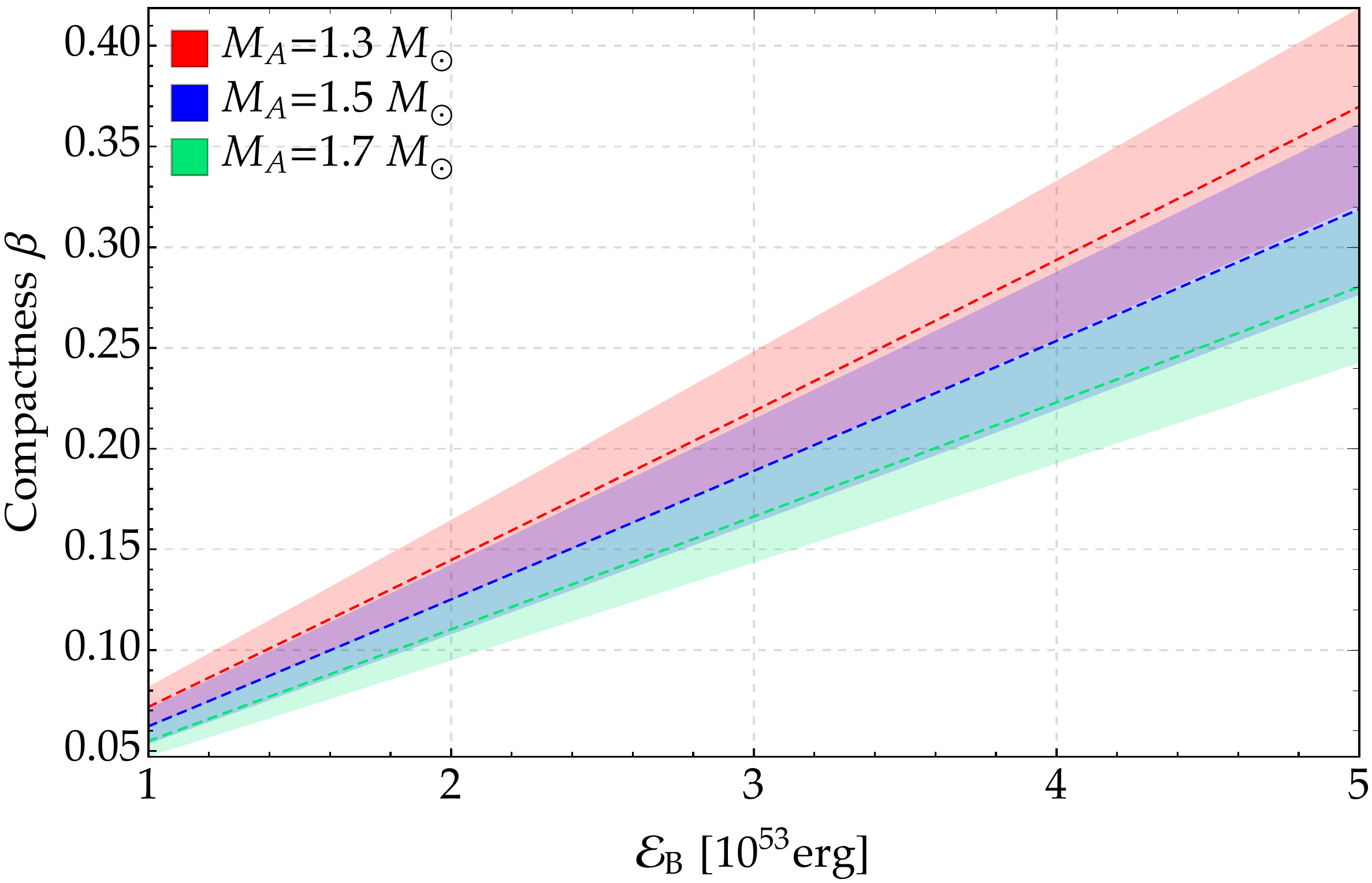}
		\caption{
		Compactness$-\mathcal{E}_{\mathrm{B}}$
		constraint}\label{fig:eb-beta}		
	\end{subfigure}
	\begin{subfigure}[t]{0.48\textwidth}
		\centering
		\includegraphics[width=\textwidth]{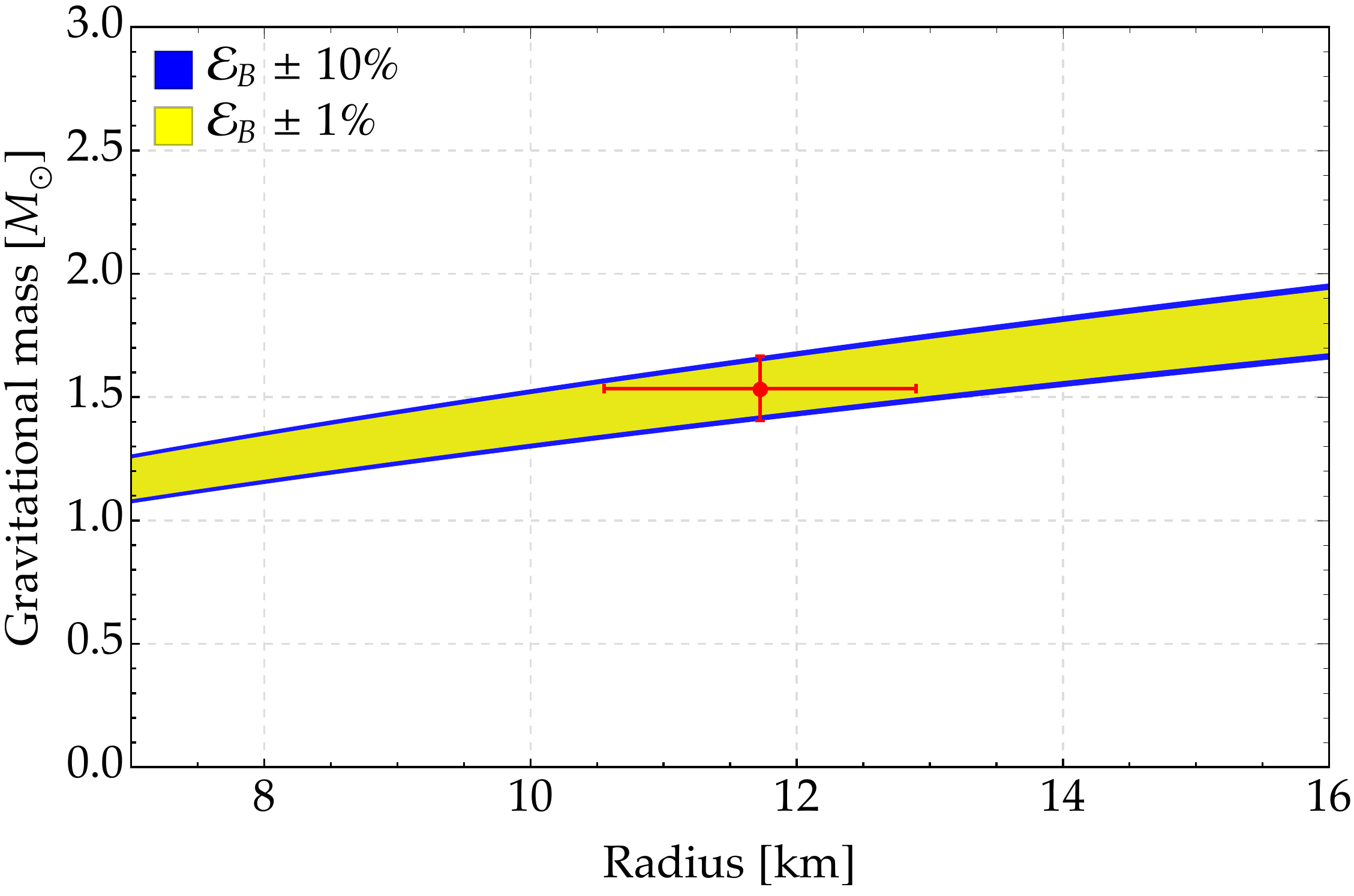}
		\caption{Mass--Radius constraint}
		\label{fig:mass-radius}
	\end{subfigure}
	\caption{Neutron-star compactness $\beta$
as a function of the gravitational binding energy
$\mathcal{E}_{\mathrm{B}}$ (\subref{fig:eb-beta}),
known with $1\%$ accuracy,
for different values of the baryonic mass $M_b$
supposed to be known with $10 \%$ precision from
the observation of \textsuperscript{56}Ni and the
pre-supernova model.  
Figure (\subref{fig:mass-radius})
presents the gravitational mass--radius allowed
region, from eq.~\eqref{e:m-r}.
The red point is a putative measurement of the
neutron-star radius and the 
corresponding neutron star mass from a fit on the
EOS given by relation \eqref{e:beta-eb}.}\label{fig:constraints}
\end{figure}

Figure~\ref{fig:eb-beta} shows the
$\beta-\mathcal{E}_{\mathrm{B}}$
relation given by eq.~\eqref{e:beta-ebb} for different values
of the baryonic mass known with $10 \%$ uncertainty.
The binding energy $\mathcal{E}_{\mathrm{B}}$ is
assumed to be determined more precisely than $M_b$
thanks to a future supernova galactic explosion.
For definiteness we assume 
$\delta\mathcal{E}_{\mathrm{B}} \sim 1 \%$.
For a nominal value $\mathcal{E}_{\mathrm{B}} \approx
3.53 \times 10^{53}$
erg the neutron-star compactness would be of
$\beta = 0.258\pm 0.035,\, 0.223\pm 0.030,
\, 0.196 \pm 0.027$
for the representative values
$M_b / M_{\odot} = 1.3,\,1.5,\,1.7$ respectively.

Figure~\ref{fig:mass-radius} presents the gravitational
mass--radius allowed region, based on relation
\eqref{e:m-r} from neutron-star EOS.
We show a blue band in the achievable case
of $10 \%$ accuracy in the binding energy measurement,
if a supernova explodes tomorrow. 
The figure also shows a yellow band corresponding to $1 \%$
precision in $\mathcal{E}_{\mathrm{B}}$.
In fact, this should be achievable in a similar
analysis as the one performed here
with operative detectors of  increased fiducial
volumes, e.g.\ with Hyper-Kamiokande, and by combining more
detection channels in various observatories. 
The red point indicates the allowed gravitational mass  of
$M = 1.535\, M_{\odot}$ for
the newly formed neutron-star from relation \eqref{e:m-r}, 
if one assumes that information on the radius is
obtained at $10 \%$ level, e.g. $R = (11.73 \pm
1.17)$ km. Notice that chiral effective field
theories give interesting theoretical constraints
on neutron star radii \cite{Hebeler:2010jx}.
Moreover, the measurement of gravitational
waves from binary
neutron star mergers, as the one recently
observed \cite{TheLIGOScientific:2017qsa},
will bring significant
constraints on the neutron stars EOS and radii.

In summary, we have shown that the gravitational
binding energy of the next galactic
supernova can be
measured through its neutrinos in a detector
such as Super-Kamiokande with a precision of
$\sim 10 \%$, without any priors
and by combining inverse beta decay
and elastic scattering events.
Increasing volumes and number
of detection channels is likely to reduce
this error greatly; e.g., the statistical improvement due to
the increased mass in Hyper-Kamiokande is a
factor of
$\approx 4$.
We have shown here that this information
can be used e.g.\ to determine the
neutron star compactness with a precision
of $\sim 10\%$; the limiting factor is the
uncertainty on the baryon mass $M_b$ and not the
one on $\mathcal{E}_{\mathrm{B}}$.
Future EOS investigations are likely to better
constraint the $\beta-\mathcal{E}_{\mathrm{B}}$
relation \eqref{e:beta-eb} allowing more precise
inferences. Clearly, a precise measurement of 
$\mathcal{E}_{\mathrm{B}}$, achievable in
future observation of supernova neutrinos,
would provide crucial observational constraints on
the supernova mechanism, on neutrino properties
and on the neutron star mass--radius relation.

\section*{Acknowledgments}
\addcontentsline{toc}{section}{Acknowledgments}

We would like to thank the precious support given by
Walter Fulgione during the preparation of this work,
as well as the kind help on neutrino-oxygen cross section
given by Karlheinz Langanke, Gabriel
Mart\'{\i}nez-Pinedo
and Andre Sieverding.
M.C.\ Volpe would like to thank Achim Schwenk and
Jim Lattimer for useful discussions. She acknowledges
support from  ``Gravitation et physique fondamentale''
(GPHYS) of the Observatoire de Paris.

\end{document}